\documentclass[np2,twoside,onecolumn,showpacs]{revtex4}

\usepackage{amssymb}
\usepackage{amsmath}
\usepackage{bm,multirow}

\def\bi{\bibitem}
\def\la{\langle}\def\ra{\rangle}\def\be{\begin{eqnarray}}\def\ee{\end{eqnarray}}
\def\roughly#1{\mathrel{\raise.3ex\hbox{$#1$\kern-.75em%
\lower1ex\hbox{$\sim$}}}}
\def\lsim{\roughly<}
\def\gsim{\roughly>}

\begin{document}

\title[Probing the Source of Proton Mass by\\  ``Unbreaking" Scale-Chiral Symmetry]{Probing the Source  of Proton Mass by\\  ``Unbreaking" Scale-Chiral Symmetry\footnote{Prepared for ``New Physics: Sae Mulli" of Korean Physical Society.} }

\author{Mannque Rho}
\email{mannque.rho@cea.fr}
\affiliation{Institut de Physique Th\'{e}orique, CEA Saclay, 91191 Gif-sur-Yvette, France}



\begin{abstract}
I describe a possible scenario for the origin of proton mass  in terms of Cheshire Cat, half-skyrmions, topology change and interplay between hidden chiral-scale symmetry and induced local symmetry. This differs from the standard constituent-quark scenario. As the baryonic matter density is increased toward the vector manifestation (VM) fixed-point at which the $\rho$ mass is to vanish, the effective in-medium mass ratio $m^*_\rho/m^*_N$ is to tend to zero proportionally to $g^*_\rho$ where $g^*_\rho$ is the in-medium hidden gauge coupling constant. I develop the thesis that the intricacy involved in the mass generation could be decoded from experiments at RIB accelerators and massive compact stars.
\end{abstract}


\maketitle

\section{Introduction}
Where the proton (or generally nucleon) mass comes from  remains, even after the discovery of Higgs boson, as one of the great mysteries of Nature. In this note I explore the possibility that the answer could perhaps be found in nuclear physics. Given that the mass of the proton  is very accurately measured, $938.272046\pm 0.000021$ MeV,   more than 99.9\% of the ``visible" mass around us -- all the way down to atoms -- can be accounted for in great accuracy by adding the number of nucleons involved in the system.  Even that of the nucleus which is in the core of atoms is accounted for up to 98\%.   The  strong interactions taking place inside the nucleus are now quantitatively described by quantum chromodynamics (QCD). So the mass of a nucleus is nearly completely given by the sum of the mass of the nucleons in the nucleus with only a small correction of binding energy, less than 1\% of the proton mass, and even that small value can be fairly well explained, though in a highly intricate way, by QCD. This additive accountability, however, ends abruptly -- and singularly -- at this point. The proton mass, the source for the nuclear mass, is no longer accounted for as a sum of ``something."  In QCD, the relevant constituents are quarks, and for the nucleon the (current) quarks involved, ``up" and ``down," are very light,  less than 5 MeV.  Thus how the proton mass arises from its constituents must be drastically different from how the mass of a nucleus comes about. Effectively, QCD for the proton is a theory for ``mass without masses"\cite{wilczek}.

For most of particle physicists, however, the proton mass is satisfactorily  ``explained."   Hence the end of the story?  In Wilczek's words\cite{wilczek2}, ``Here there are no uncontrolled approximations, no perturbative theory, no cutoff, and no room for fudge factors.  A handful of parameters, inserted into a highly constrained theory of extraordinary symmetry, either will or won't account for the incredible wealth of measured phenomena in the strong interactions. And it certainly appears that they do." Indeed lattice simulation with QCD  with no mass terms and with the heavy quarks $c$, $b$ and $t$ ignored, termed ``QCDLite" by Wilczek,  predicts  $\sim 95\%$ of the proton mass. If pressed to explain in more detail, they will then say it comes from the massless gluons interacting with the nearly massless, high-momentum quarks, winding up confined in a ``bag."   In a nutshell, one may say the mass is generated due to the breaking of chiral symmetry associated with the nearly massless quarks by the confinement\cite{casher-conf}.  In the standard paradigm, the chiral symmetry spontaneously broken leading to the mass generation is characterized by the quark condensate, $\langle\bar{q}q\rangle\neq 0$, the order parameter for chiral symmetry.

To nuclear physicists, ironically, this explanation raises more questions than answers. This may be due to the fact that confinement is very poorly -- if at all -- understood. For mathematicians, it is ``One-Million Dollar Clay Millenium Problem," still to be solved. It may even be, as some argue, that ``deconfinement," corollary to confinement, does not take place within the framework of QCD\cite{glozman}.  For nuclear physicists,  it presents numerous puzzles, some of which bear directly on observables as I will describe below.

\section{The Cheshire Cat}
\subsection{Chiral bag and confinement}\label{CC}

One extremely simple picture of how confinement could lead to the spontaneous breaking of chiral symmetry is to look at a massless Dirac particle -- let me call it quark -- swimming in one space dimension to the right on top of the Dirac sea. Suppose there is an infinitely tall impenetrable wall on the right of what I will call the ``jain cell."   The swimmer, bumping into the wall, has no choice but to turn back or drown. Now chiral symmetry forbids the swimmer from turning and swimming back unscathed to the left on {\it  top} of the Dirac sea.  It  could swim back on top of the sea only  if the quark picked up a mass, that is, if chiral symmetry were broken -- explicitly -- by the wall. The wall of the ``jail-cell" could then be the source of the mass\cite{CC}. This is the well-known MIT bag picture of the hadron. But how does this give the mass? The jail wall, one might say, is erected by the as-yet un-understood strong quark-gluon dynamics and the existence of the mass is then the consequence, in the QCDLite, of the chiral symmetry broken by the jail wall, making the quark condensate nonzero. This may sound fine, but there is something missing here, and that is the Nambu-Goldstone theorem that chiral symmetry realized with a non-zero condensate in massless theory like QCDLite must have zero-mass pions, namely,  Nambu-Goldstone (NG) bosons.

It has been understood since many years that pions are absolutely essential for nuclear physics, although it was not clear how they figured precisely until meson-exchange currents were understood with the advent of chiral perturbation theory.  There are many different ways to bring in pions and have them couple to the quarks in the nucleon. This accounts for the variety of models found in the literature that incorporate chiral symmetry, such as chiral bag, chiral soliton, cloudy bag etc. It is very possible that different ways, constructed in consistency with symmetries and invariances involved, could lead to qualitatively similar results.  Here I will adopt the approach called ``chiral bag" since it is in that picture that I can address the issues involved with some confidence.

In the chiral bag model\cite{chiralbag}, both the pions  and the confinement of quarks/gluons are taken to be indispensable in the structure of the elementary proton as well as that of the nucleus. In this model, the pion, a NB boson (in QCDLite), is not just a propagating field but contains highly nonlinear mean-field components that encode topology. Let us then look again, this time taking into account the intricacy involving the NG boson, at the one-dimensional swimmer confined in the jail-cell\cite{CC}. As the quark $\psi$ bumps into the wall, it is subject to a boundary condition imposed by the wall, $in^\mu\gamma_\mu\psi=-\psi$ where $n_\mu$ is the outwardly-directed unit 2-vector and $\psi$ is the quark field. This is a classical condition imposed by the vector current conservation. One can easily see that $\bar{\psi}in^\mu\gamma_\mu\gamma_\mu\psi=0$, so the quark charge is confined. But we are forgetting here that there is pion that couples, due to chiral invariance, to the quark at the surface. The coupling modifies the boundary condition to
 \be
in^\mu\gamma_\mu\psi=-e^{i\gamma_5\phi/f}\psi
\ee
where $\phi$ is the one-dimensional ``pion" field and $f$ is a constant -- one dimensional analog to the pion decay constant $f_\pi$. Classically the charge is still conserved. But quantum mechanically, the coupling brings in divergences, and as is well known in field theory the divergence has to be regulated to make the theory sensible. The regularization turns out make the fermion (quark) charge $Q$ no longer conserved, making it leak out of the bag as  $\frac{d}{dt} Q(t)=\frac{1}{2\pi f} \phi(t)$. Formally this is a consequence of the vector-current anomaly generated at the bag boundary when the axial current conservation is imposed.  For the quark swimmer, what happens at the bag wall then is that the helicity conservation forces the swimmer to plunge into the Dirac sea and swim back to the left under the Dirac surface. 

This plunge into the occupied Dirac sea, made possible due to what is called ``infinite-hotel phenomenon"\cite{infinitehotel}, however, makes the fermion charge disappear from the system. But we have a theory in which the charge should be conserved. So what happens to the fermion charge that disappears as the quark plunges into the sea? 

Here the penny drops!  The leaking charge is picked up by the soliton from the pion outside of the bag.  In terms of the so-called ``chiral angle" $\theta (R) = \phi(R)/f$ where $R$ is the position of the bag wall, the leaking charge depends on the size of the cell,  $\Delta Q (R) = \theta (R) /\pi$ . The soliton carries away precisely this leaked charge. For what's referred to as the ``magic angle" $\theta_m=\pi/2$, the fermion charge is precisely 1/2 inside and 1/2 outside of the bag. For any angle $\leq \pi$, it is partitioned in such a way that the total is precisely 1.

Involving topology, exactly the same leakage of charge should take place also in (3+1) dimensions and indeed it does. The soliton that carries the leaking charge is the famous skyrmion\cite{skyrme}. This was first shown in \cite{rho-goldhaber-brown} for the case of  the ``magic angle" in $(3+1)$ dimensions where the baryon charge is partitioned half-and-half as in $(1+1)$ dimensions. It turns out, however, that even though many nuclear properties seem to work well when described at the ``magic angle," there is nothing special -- except for numerical efficiency -- about this angle. The partition does indeed take place for {\it any} chiral angle as was shown quite convincingly by Goldstone and Jaffe\cite{goldstone-jaffe}.   In this picture, the fermion (baryon) charge is {\it not confined} inside the bag.  This phenomenon is dubbed ``Cheshire Cat phenomenon"\cite{CC} drawing an analogy from the famous smile of Cheshire Cat in ``Alice in the Wonderland" of Lewis Carol.

One might attribute the above CC phenomenon to the topological quantity.  What about processes that are not topological? What about, for instance, the color charge that has to do with confinement but is not connected to topology? Does the color charge leak?

The answer is: Yes it also leaks. This is because just like the baryon charge, the color charge is also broken by the vector anomaly. Classically the color is confined within the bag by the vector current conservation, but quantum mechanically it leaks out at the bag boundary. Here the culprit is the $\eta^\prime$ associated with the $U_A(1)$ anomaly. Lodged outside of the bag, like pions, the $\eta^\prime$ couples to the quark at the boundary and induces a color electric field normal to the surface.  But unlike the baryon charge which can be transferred to the soliton in the pion outside, there is nothing outside that can carry the leaking color charge. The $\eta^\prime$ causing for the anomaly is color singlet. The only way color gauge symmetry can be preserved is then to append to to the wall a color-symmetry-breaking counter term  to stop the flow of quantum-induced color\cite{colorsymmetry}. Note that this counter term is classical, so the Lagrangian at the classical level is color-gauge non-invariant. This is contrary to what usually happens in gauge theories without boundaries where symmetries conserved in the Lagrangian are broken quantum mechanically, such as , e.g., $U_A(1)$ anomaly, trace anomaly etc..

\begin{figure}[ht]%
\includegraphics[width=10cm]{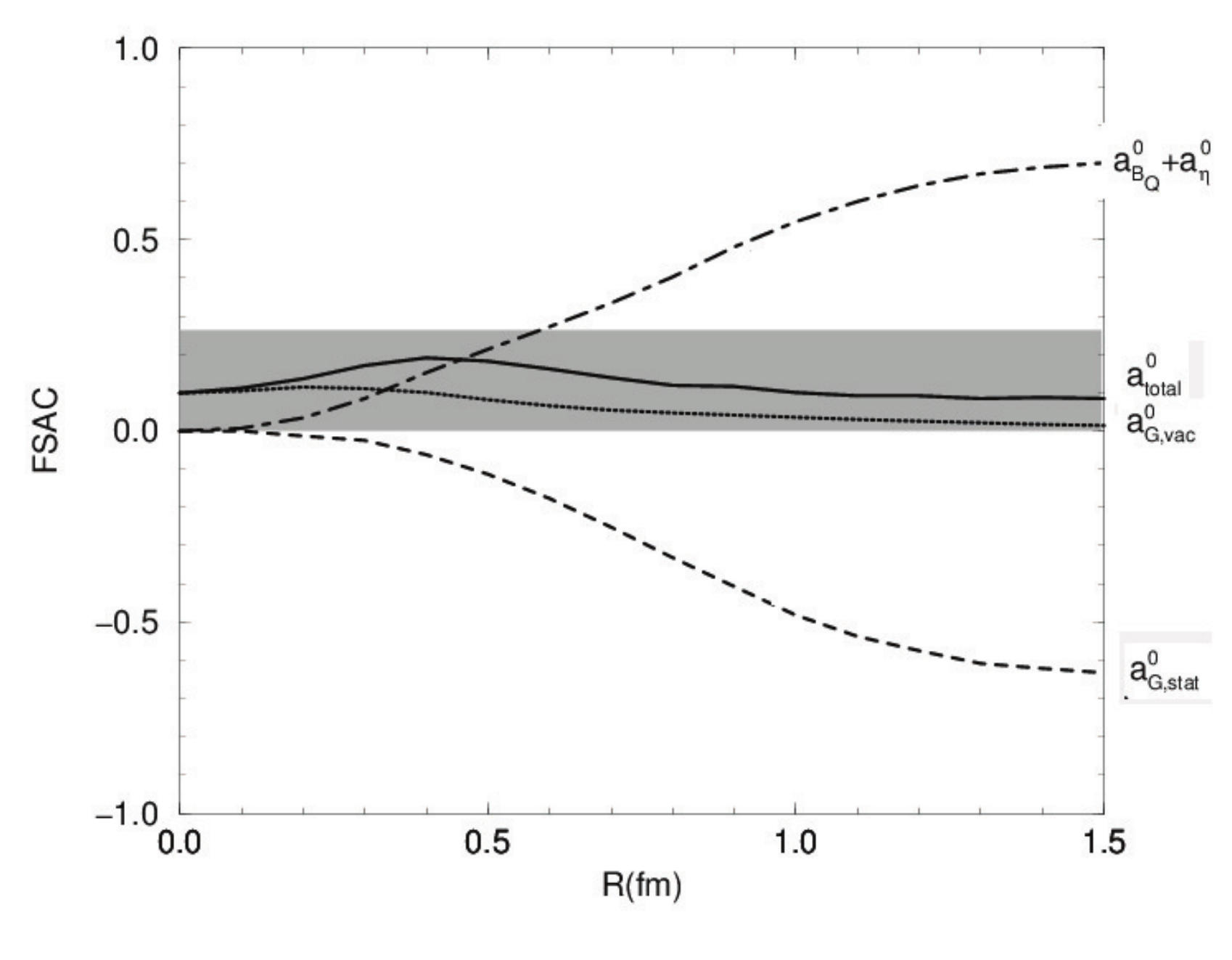}
\caption{Various contributions to the FSAC in the chiral bag model: : (a) quark
plus $\eta^\prime$ (or ``matter") contribution ($a^0_{B_Q} + a^0_\eta$),
(b) the contribution of the static gluons due to quark source
($a^0_{G,stat}$), (c) the gluon Casimir contribution
($a^0_{G,vac}$), and (d) the sum of (a)-(c) ($a^0_{total})$. The shaded area
corresponds to the range admitted by experiments.}\label{FSAC}
\end{figure}
\subsection{Flavor singlet axial charge of the proton}

How does the Cheshire Cat phenomenon manifest itself in physical processes that involve neither baryon charge (topology) nor color charge (gauge invariance) directly? A good example for answering this question is the flavor singlet axial coupling constant (FSAC)  $g_A^{0}$ in the current $J_{5 \mu}^0=\frac 12 g_A^{0}\bar{\psi}\gamma_\mu\gamma_5\psi$. The FSAC became famous because of the so-called ``proton spin crisis" which turned out be a false alarm. The $g_A^{0}$ in general is a form factor depending on the momentum transfer involved which in some kinematical condition is  related to the proton spin. But the axial charge at zero momentum transfer that I will denote as $a^0=g_A^0 (0)$ has nothing directly to do with the proton spin. What is of relevance to what I am discussing here intricately involves quarks and gluons inside the bag and $\pi$ and $\eta^\prime$ outside -- with the baryon charge fractionization and the color-charge anomaly cancelation -- so in the chiral bag model topology as well as confinement figure in $a_0$.  One can write down a simple chiral-bag Lagrangian that encodes all the above properties.  One was written down in \cite{colorsymmetry} and applied to the explicit calculation of the FSAC in \cite{fsac-cc}. The details are given in the references, so I won't go into them. The  result, shown in Fig.~\ref{FSAC}, illustrates clearly how the Cheshire Cat principle works.  While the individual contributions from inside and outside are widely varying in terms of the  ``confinement radius"  $R$ the total has, within the approximations made, no visible dependence on $R$.  The bag model (e.g., ``MIT bag") and the skyrmion are in a sense equivalent. This is a signal that at low energy/momentum, physical properties may very well be ``blind" to the ``confinement" size. In fact, one can even formulate, within the framework of chiral bag, the  notion that the ``bag size"  is a gauge artifact, meaning that one can pick any $R$ without changing physics\cite{BR:DD}. In this description, deconfinement seems irrelevant.

So what can one say about the source for the proton mass based on the Cheshire Cat phenomenon? Up to here, it only connects the skyrmion to the bag but says nothing on the mass. We will come  to this matter below.

\section{The Proton in Large $N_c$ QCD}
According to a general argument for baryon structure in QCD deduced in the large $N_c$ limit, the only nonpertubative tool available for approaching QCD in nonperturbative regime of high density, the constituent quark model should be equivalent to the skyrmion model. I will exploit this equivalence in what follows.
\subsection{Constituent-quark (quasiquark) model}
In Nambu's pathbreaking paper\cite{nambu-jona-lasinio}, it is stated in conclusion ``... Our model Hamiltonian, though very simple, has been found to produce results which strongly simulate the general characteristics of real nucleons and mesons. It is quite appealing that both the nucleon mass and the pseudo-scalar `pion'  are of the same dynamical origin, and the reason behind this can be easily understood in terms of (1) classical concepts such as attraction or repulsion between particles, and (2) the $\gamma_5$  symmetry."  For convenience I will refer to this statement as   defining ``Nambu mechanism."\footnote{Of course, Nambu's idea of mass-generation as the spontaneous breaking of global symmetry is a lot more general than just the proton mass and the pion. It embodies the whole  range of ``spontaneous symmetry breaking (SSB)"  in science in general, not just in particle physics such as the Standard Model Higgs boson, but also in condensed matter.}

In the context of modern development, it seems more appropriate to replace the ``proton"  in the Nambu scenario by ``constituent quark" -- that I shall equivalently refer to as ``quasiquark" (denoted $Q$) -- and to view the proton (nucleon) as a bound state of three $Q$s with the mass $m_Q\sim \frac 13 m_N$ generated by the Nambu mechanism. In this model a meson, say, $\rho$ meson, is  a constituent-quark-constituent-antiquark ($Q\bar{Q}$) bound state.  The interactions between $Q$s are supposed to be weak, thereby justifying the term ``quasiquark" in analogy to quasiparticles in many-nucleon systems.  This $Q$ picture is justifiable in the large $N_c$ limit from QCD\cite{weinberg-constituent}. In fact, this description is known to work very well, encompassing all light-quark hadrons, mesons and baryons.  If the quasiquark mass $m_Q$ arises from the Nambu mechanism of chiral symmetry breaking (CCB),  then if one were to ``unbreak"\footnote{I am borrowing this terminology from S. Weinberg\cite{weinberg-unbreak}.} the symmetry by density (or temperature) such that $\la\bar{q}q\ra\to 0$,  then we should expect that $m_Q\to 0$.
\vskip 0.3cm
$\bullet$ {\bf Q(uasiquark) Scenario}:  In approaching  $n_c$ at which $\la\bar{q}q\ra\to 0$,  the masses would scale $m_N^*\to 0$ and $m_\rho^*\to 0$ in such a way that ${\cal M}\equiv m_\rho^*/m_N^*\to \sim 2/3$. Here it is assumed that  the quasiparticle picture holds in dense matter.

\subsection{Skyrmions and half-skyrmions}
The sharp bag boundary set up in the discussions given above could very well be an artifact of approximation. In Nature, the leakage process could be smooth. The Cheshire Cat implies that when the bag is shrunk, all properties of quarks and gluons will be transferred to the topological soliton, the skyrmion. The mass generated in the  QCDLite by confinement must then be lodged in the lump of soliton. In this picture there is nothing special with single skyrmion, proton. It can be a nucleus with the mass number $A$ given by the winding number $A$. In the mathematics community, this feature is being exploited to unify in terms of complex geometry the proton structure to the nuclear as well as atomic structure\cite{geometry}. Seen from this vista, there is nothing special about the proton from, say, the carbon-12 nucleus.

Indeed, the nucleus $^{12}$C including the subtle structure of Hoyle states can be understood as a skyrmion with winding number 12\cite{lau-manton}. It is equally well described in chiral perturbation theory with baryonic chiral Lagrangians\cite{c12-chpt} and also in no-core shell model\cite{nocore}. The dynamics involved in these descriptions, ranging from skyrmions to effective field theories  \`{a} la Weinberg's folk theorem to the traditional shell-model implemented with symmetry properties of chiral symmetry seem to tell more or less the same story.  The key question is: How can one decode the source of the ``proton mass" going from skyrmions to quark-gluon structure as indicated by the Cheshire Cat phenomenon which is most likely different form the $Q$ scenario?

To address this question, let us ``unbreak" the symmetries involved -- e.g.,  ``uncondense" the quark condensate -- by going to highly dense baryonic matter as one expects to encounter in compact stars. To do this, I consider putting skyrmions on crystal lattice and squeezing the lattice size.

If one accepts that (1) the skyrmion structure is a good description of nucleons in the large $N_c$ limit and (2) skyrmions on crystals are a reliable model for baryonic matter at high density,  then the first observation is the remarkable phenomenon that as density is increased beyond the normal nuclear matter density $n_0\sim 0.16$ fm$^{-3}$, the skyrmions in baryonic matter  fractionize into half-skyrmions. At present, due to incomplete understanding of the model, it is not feasible to give an accurate value for the transition density, but based on certain sophisticated Lagrangians to be described below combined with available phenomenology,  the transition density is estimated to be $\sim 2n_0$. It cannot be much lower as there is no indication from available experimental data. It cannot be much higher, for in that case, the model cannot be trusted. This is just about the density regime that one expects to be able to probe in various accelerators for nuclear physics, for example, RAON in Korea. Also that a soliton can turn into various fractionized solitons is widely conjectured and in some cases observed in Nature (see for instance \cite{multifacet}). For the skyrmion in question, one can actually mathematically verify that the fractionization of a skyrmion takes place in the presence of a certain potential term of ``Heisenberg type"\cite{fractionization}. Such a potential is not present in (generalized) Lagrangians currently studied so it is somewhat academic, but as I will describe there is scale symmetry that could generate such a potential in dense medium\cite{LPR}. In any event, there is an indirect indication for half-skyrmion structure already in the $\alpha$ particle in which four nucleons are strongly bound\cite{BMS}.

The fractionization on crystal lattice at a density $n_{1/2} \gsim 2n_0$ has been observed since some time in the Skyrme model (with pions only) or generalized models with massive degrees of freedom \cite{Park-Vento,HLMR-multifacet}. As indicated in the mathematical investigation\cite{fractionization}, the fractionized skyrmions are most likely confined, non-propagating degrees of freedom.  Of great significance to what I am discussing here is that when the skyrmions fractionize into half-skyrmions, the quark condensate $\la\bar{q}q\ra$, which is non-zero both globally and locally in the skyrmion phase, vanishes when averaged over the crystal lattice. Let me denote this averaged condensate by $\Sigma\equiv \overline{\la\bar{q}q\ra}$. While vanishing on the average, however, the chiral condensate $\la\bar{q}q\ra$ supports chiral density wave in the half-skyrmion phase\cite{density-wave}.  This means that there is a sort of phase transition in the change of topology with the condensate vanishing on averaging.  But there is still pion in the matter, with nonzero $f_\pi$, hence chiral symmetry is still broken. Note that there is no obvious local-field order parameter. It is an unusual state of matter, but such states are observed in condensed matter physics\cite{multifacet}.

The questions then are: Is this phenomenon, if not an artifact of crystal structure,  relevant to nuclear physics? If it is, how to implement this phenomenon, a``topology change," in the proton mass problem? The answer to the first question is yes and the answer to the second is what I turn to next.

\section{Hidden Symmetries}
The first question raised above is generic and does not depend on what other degrees of freedom than pion are involved. However the answer to the second question involves both vector mesons and scalar mesons. They influence neither the topological structure of matter -- which is controlled by the pion field only--  nor the Cheshire Cat phenomenon. But it is found that they play extremely important roles for nuclear interactions within the framework of the effective field theory model I am using. Two hidden symmetries are found to be involved: One, hidden gauge symmetry for the vector mesons $V=(\rho,\omega)$ and the other,  hidden scale symmetry for a low-mass scalar meson ($\sigma$). From here on I will focus on the $\rho$ meson (the hidden symmetry argument applies also to $\omega$, though in a different way in dense medium) and the dilaton $\sigma$ associated with the spontaneous breaking of scale symmetry.
\subsection{Hidden local symmetry}
Nuclear interactions that enter in nuclear structure probe energy/momentum scales up to order of the vector ($\rho$)-meson mass. In the spirit of effective field theory, a natural procedure is then to introduce the $\rho$ field explicitly.  A powerful -- and perhaps the most predictive -- way is to resort to the hidden local symmetric Lagrangian which is gauge-equivalent to non-linear $\sigma$ model\cite{HLS}. The pion field of the sigma model $U=e^{i\pi/f_\pi}=\xi_L^\dagger\xi_R \in (SU(N_f)_L\times SU(N_f)_R)/SU(N_f)_{L+R}$ has the redundancy $\xi_{L,R}\to h(x)\xi_{L,R}$ with $h(x)\in SU(N_f)_{L+R}$.  This redundancy can be exploited to elevate the energy/momentum scale to the vector meson mass scale by gauging the vector $\rho$. This leads to hidden local symmetry with the kinetic energy generated dynamically by loop corrections\cite{HLS}. There is magic in this approach if one assumes that the vector meson mass can be taken ``light"\cite{HLS,komargodski}. It is not understood how the magic comes about, but when treated to the leading order (at which calculations can be done easily) it gives some remarkable results such as the KSRF relation (to all-loop order), vector dominance etc. which are difficult to obtain in standard chiral perturbation approaches. Associated with the magic that goes with ``light" vector meson, there is one specially distinctive prediction which is not present in other approaches, namely,   at some high density (and perhaps also at high temperature) which may precede possible deconfinement, the $\rho$-meson mass approaches zero as the critical density (to be denoted $n_{VM}$) is approached. A renormalization-group analysis\cite{HLS} showed that as the quark condensate is dialed to zero, the $\rho$ mass which is given by the KSRF formula $m_\rho^2=a f_\pi^2 g_\rho^2$ goes as
\be
m_\rho\sim g_\rho\to 0.
\ee
It should be noted  that it is not the pion decay constant but the hidden gauge coupling $g_\rho$ that controls the property of the $\rho$ mass. This can be understood with hidden scale symmetry (discussed below):. The $\rho$ mass is scale-invariant in the vacuum and is expected to remain scale-invariant in medium.
The point at which the mass vanishes is referred to as ``vector manifestation (VM) fixed point." As stressed in \cite{HLS}, the VM fixed point is {\it not} in QCD proper. It may be not only {\it hidden} but also {\it non-reachable} in QCD in the vacuum. However it is suggested that it can be generated in highly dense medium. It will turn out in what follows that the approach near to the VMFP -- which may be different from the chiral symmetry restoration density $n_c$ -- can play a crucial role in compact-star matter .

\subsection{Hidden scale symmetry}
Another hidden symmetry highly relevant to nuclear dynamics is global scale symmetry, presumably connected to a low-mass scalar $\sigma$. QCD at classical level is scale-invariant in ``QCDLite," i.e., $m_q\to 0$ for $q=u, d$ for non-strange nuclear interactions and also $s$ if strangeness is involved, but the scale symmetry is broken quantum mechanically by the QCD trace-anomaly. It is renormalization-group invariant, so cannot be turned off in QCD dynamics. However similarly to the $U_A (1)$ anomaly, it may be restored or rather made ``emerging" in medium, particularly at high density. It is this  possibility that I will explore for probing the source of the proton mass.

A scalar meson of mass $\sim 500$ MeV has been playing a very important role in nuclear physics, in, say, nuclear forces, relativistic mean-field models etc. But it has been totally unclear what that scalar is in QCD\cite{pelaez}. In particle physics, the ``low-lying" scalar Higgs scalar $H$ poses a similar puzzle. In one of the candidate scenarios for Higgs, it is identified as a NG boson associated with broken scale symmetry with an infrared (IR) fixed point for $N_f\sim 8$\cite{yamawaki-GEB}.

Although difficult to identify, it is surprisingly simple to see where that scalar is ``hiding" in the strong interactions. The simplest description was given by Yamawaki for Higgs physics and going beyond the Standard Model\cite{yamawaki-GEB}. The reasoning made in the strong interactions is quite similar as the SM Higgs Lagrangian is known to be equivalent to this linear sigma model.  

Let us start with the linear sigma model for flavor $SU(2)$ with the quartet of scalars, i.e., the triplet of pions $\pi$ and a scalar $\phi$.  (I shall avoid the notation $\sigma$ customarily used in the literature for linear sigma model. It is reserved for the dilaton.) There is one parameter, $\lambda$, in the model which can be dialed to strong coupling limit ($\lambda=\infty$) or to weak coupling limit ($\lambda=0$). In the strong coupling limit, one obtains the usual nonlinear sigma  model with the scalar $\phi$ absent, hence no scale symmetry. The usual current algebra term embodying low-energy strong interactions is of course not scale-invariant. Standard chiral perturbation theory with baryons explicitly implemented is based on this framework. On the other hand, the weak coupling limit $\lambda=0$  gives rise to a scale-invariant nonlinear sigma model with scale symmetry broken only in a potential that encodes the trace anomaly of QCD. In the dilatonic Higgs model of \cite{yamawaki-GEB}, the Higgs mass can be small because it is near the IR fixed point.

Now the central  point of this article: {\it In nuclei and nuclear matter, by dialing density, one can sample from the strong coupling regime to the weak coupling regime.} In the strong coupling limit which is applicable to low-energy hadronic interactions in the matter-free vacuum, nonlinear sigma model works, with the fourth component of the chiral four vector banished to infinity (say $> m_N\sim 1$ GeV).  However in nuclei and nuclear matter at $n\sim n_0$, one knows from Walecka-type relativistic mean-field (RMF) model that a scalar of $\sim 600$ MeV is needed, supplying the attraction that binds nucleons. What is needed here is not $\phi$ that figures in the Gell-Mann-Oakes-Renner (GMOR) linear sigma model but a chiral singlet with little mixing into the fourth component field.  Near nuclear matter density\footnote{At low density below $n_0$, there is gas-liquid phase transition. I won't deal with it,  focusing entirely on $n\gsim n_0$.}, the relevant degrees of freedom are nucleons, a scalar of $\sim 600$ MeV and the vector mesons $\rho$ and $\omega$ of $\sim 700$ MeV treated at the mean field (with pions entering at loop order) giving  rise to relativistic mean-field theory (RMFT) which is roughly equivalent to Landau Fermi liquid theory. Here there is no scalar corresponding to the fourth component of the chiral four-vector. As density is increased above $n_0$, however, it is expected that the pion and a scalar will start to approach closer to the structure of linear sigma model, ultimately forming the massless quartet (in the chiral limit). Such a scenario was first proposed by Beane and van Kolck\cite{beane-vankolck}, and then given a support with a scale-invariant hidden local symmetry model\cite{paengetal}. The limit at which the triplet of pions and scalar form the degenerate massless quartet is referred to as  ``dilaton limit fixed point (DLFP)." There is nothing to suggest that the DLFP coincide with the VM fixed point.

One possibility\cite{LPR} that has been recently studied is that  the DLFP is connected to an infrared fixed point (IRFP) in QCD that was proposed by Crewther and Tunstall (referred to as CT)\cite{CT} and Golterman and Shamir\cite{GS}. Both postulate the presence of an IR fixed point at which scale (conformal) invariance is restored and its spontaneous breaking gives rise to a pseudo-Nambu-Goldstone boson (in the presence of the trace anomaly). The characteristic of the IR fixed point is different in the two approaches, but at the leading order of the combined scale symmetry and chiral symmetry, termed scale-chiral symmetry, they are quite similar\cite{LMR}. Here I will resort to CT.

There is a great deal of controversy as to whether such an IR fixed point exists in QCD for $N_f \sim 3$ that concerns us in nuclear physics. Some of this matter are discussed in the references cited above, including \cite{LMR}. The point I will make here is that regardless of whether such an IR fixed point exists in the chiral-scale limit in QCD, scale symmetry is relevant in nuclear physics and connected to the possible source of proton mass, can be probed in nuclear physics.
\section{Dense Baryonic Matter}
I now turn to applying the theoretical framework developed above  to nuclear phenomena. I will take the effective Lagrangian that I will refer to as ``scale-invariant HLS" that I will consider first without explicit nucleon fields ($s$HLS for short) and then with nucleon fields ($bs$HLS). 
\subsection{Consequences of topology change}
\subsubsection{\it Cusp in the symmetry energy}
\begin{figure}[ht]
\begin{center}
\includegraphics[height=5cm]{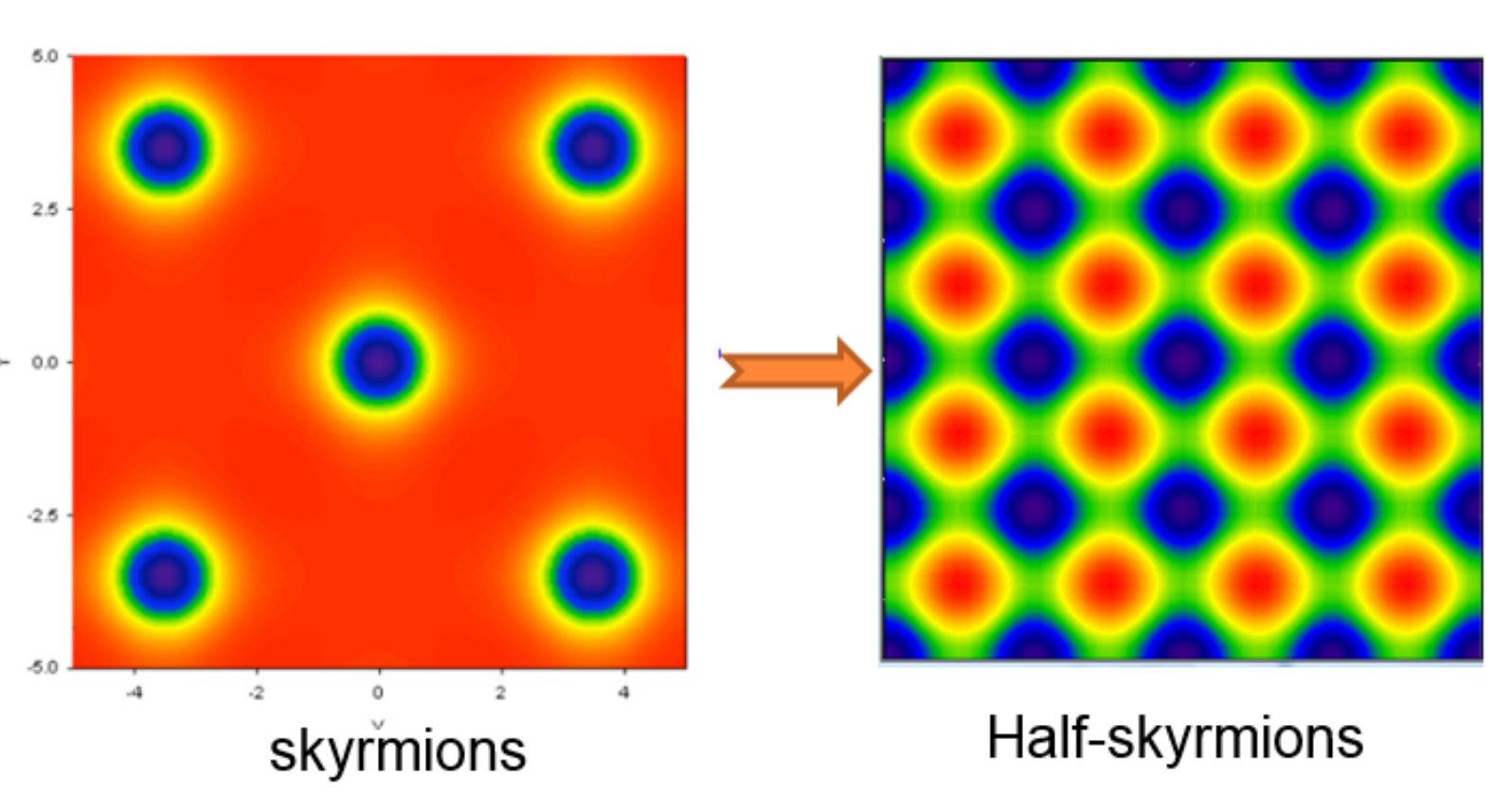}\includegraphics[height=5cm]{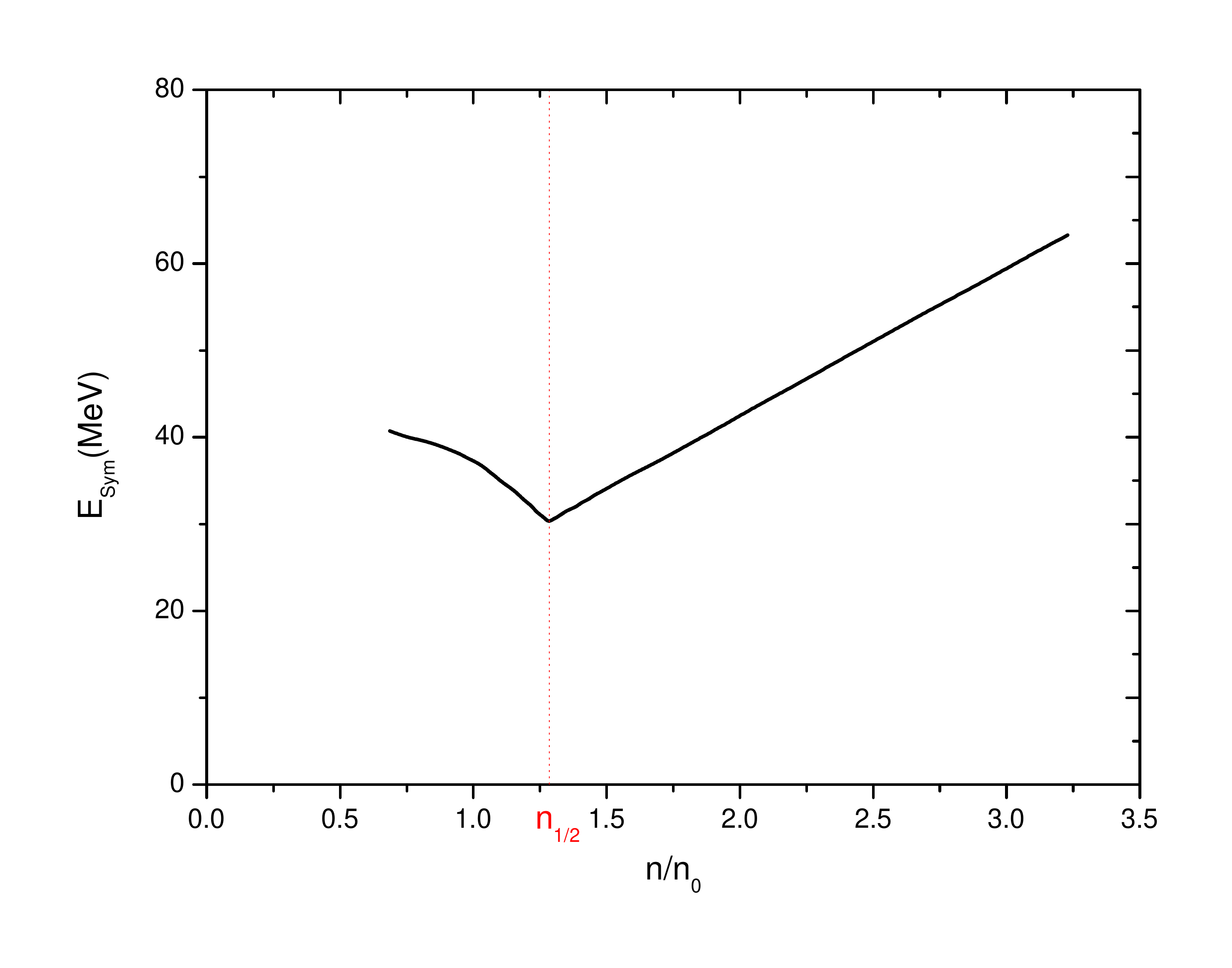}
\caption{Transition from skyrmions to half-skyrmions \cite{Park-Vento} at increasing density which leads to the cusp in the symmetry energy $E_{sym}$ at $n_{1/2}$\cite{LPR-cusp}. }
\label{cusp}
\end{center}
\end{figure}

When skyrmions are put on crystal lattice, one striking phenomenon observed is a cusp in the symmetry energy $S\equiv E_{sym}$ defined in the energy per particle $E$ of baryonic matter\cite{LPR-cusp}
\be
E(n,x)=E(n,0)+Sx^2 +\cdot\label{S}
\ee
where $x=(N-P)/A$ with $A=N+P$ and $n$ is the baryon number density. As shown in Fig.\ref{cusp}, as  skyrmions fractionize into half-skyrmions  at $n=n_{1/2}$, the cusp (Fig.\ref{cusp} (right panel)) is formed in such a way that the $S$ first decreases and then goes up. This cusp is a generic phenomenon involving the pion field only associated with the topology change.  The density at which the cusp is located, however, is determined by the other degrees of freedom than the pion figuring in the Lagrangian. While the precise value for $n_{1/2}$ is difficult to determine, with the full $s$HLS Lagrangian, it is reasonably located at $n_{1/2}\sim 2n_0$. I will take this value in what follows. The numerical details are not very sensitive to the exact location as long as it it near $2n_0$.

What produces this cusp on crystal lattice is the topology change from skyrmions to half-skyrmions at $n_{1/2}$ with the vanishing chiral condensate $\Sigma$ in the half-skyrmion phase. I will discuss below what this implies in the nuclear tensor force as the density $\sim n_{1/2}$ is approached from below.
\subsubsection{\it Parity doubling}
Another surprising observation at $n_{1/2}$ is that the effective mass of the nucleon in the skyrmion medium reveals a hidden parity-doublet symmetry with a  nonzero mass $m_0$ as $\Sigma\to 0$\cite{maetal}
\be
m_N =m_0 +\Delta m(\Sigma)\label{mass}
\ee
with $\Delta m\to 0$ as $\Sigma\to 0$ at $n\to n_{1/2}$. Since pions are still present in the half-skyrmion phase, $\Sigma\to 0$ may not be interpreted as the restoration of chiral symmetry. Nonetheless $m_0$ can be considered as a component of mass that is chirally invariant. This indicates that parity doubling in the baryon structure is ``generated" by the topology change. In fact this can be matched to a chirally-invariant mass in the parity-doubled nucleon model where $m_0$ is put {\it ab initio} into the Lagrangian as was originally done by DeTar and Kunihiro\cite{detar-kunihiro}. When the $s$HLS Lagrangian that gives rise to the mass formula (\ref{mass}) on crystal lattice is coupled scale-chiral invariantly to parity-doubled nucleons with a chirally invariant mass $m_0^\prime$ in a continuum $bs$HLS, one obtains the same mass formula as (\ref{mass}) in dense medium\cite{bshls-paritydoubled,MKH} with $m_0^\prime$ replacing $m_0$ in (\ref{mass}) and $\Delta m^\prime (\la\bar{q}q\ra)$ replacing $\Delta m(\Sigma)$.  In this continuum model, $\Sigma\to 0$ is replaced by $\la\bar{q}q\ra\to 0$, i.e., chiral symmetry is restored, with the baryons parity-doubled. In both the skyrmion crystal and the baryon parity-doubled model, $m_0\approx m_0^\prime \propto \la\chi\ra=f_\sigma$ which stays constant for $n\geq n_{1/2}$ for the former and for $n\geq n_p$ for the latter where $n_p$ is the density at which the parity-doubling takes place. It is likely that $n_{1/2}\approx n_p$.

What is significant for my purpose is that $m_0 (m_0^\prime)$ could be large, say, $\sim (0.7-0.9)m_N$.  This is not directly associated with the Nambu mechanism anchored on the spontaneous breaking of chiral symmetry. The mass $m_0 (m_0^\prime)$ is a chiral scalar. This is apparently unrelated to the $m_Q$ generated by the Nambu mechanism. So where does this chiral scalar mass come from? How is this related to the mass generated by the Cheshire Cat mechanism I started with? This is the question nuclear physicists would like to find the answer to.

\subsubsection{\it Skyrmion model vs. quasiquark model}
As noted above, the skyrmion model and then quasiquark  model for the nucleon are equivalent in the large $N_c$ limit. The quasiquark model unifies baryons and mesons and works very well when corrected with certain $1/N_c$ corrections for various static and dynamic properties. The origin of mass for both the proton and the $\rho$ is the constituent quark mass given by the dynamically generated quark condensate.  Especially, their masses have a simple ratio, e.g., $m_\rho/m_N\approx 2/3$.

Now the skyrmion approach is also a unified approach as Skyrme suggested in his seminal paper -- and more generally in $bs$HLS -- in that it involves one ``unified" Lagrangian that accounts for both mesons and baryons.  But there is no equivalent to constituent quark in the skyrmion description. Attempts to obtain a constituent $1/N_c$ skyrmion, called ``qualiton"\cite{qualiton}, have thus far been unsuccessful\cite{no-qualiton}. There are therefore no such simple mass relations between the mesons and baryons as in the quasiquark model. In fact I see no logical connection between the two in the Cheshire Cat scenario as will be specified below.

Furthermore a  tension arises between the skyrmion picture and the quasiquark  picture when we go to dense matter. While the proton mass moves to a constant $m_0\sim O(m_N)$ after shedding off $\Delta m$ subject to the decrease in the quark condensate, the $\rho$ mass which is given in medium by $m_\rho^*\approx f_\pi^* g_\rho^*$ goes to zero because $g_\rho^*$ goes to zero at the VM fixed point. $g_\rho^*$ is not directly related to the quark condensate. This dichotomy would imply the scenario for the mass as given here is drastically different from the {\bf Q} mechanism in light quark hadrons.  If one were to endow a chiral-invariant mass $m_0/3$ to a $Q$ in the proton as one would do for the parity-doubling in the $Q$ model, it would preserve the simple ratio $m_B/m_M\approx 3/2$ and other constituent-quark properties in medium as it would in the vacuum. However it would be at odds with the VM for the vector meson. Furthermore in this picture,  there would be a tremendous increase in excitations of both parities in the spectra of both baryons and mesons in medium, even at normal matter density which is accessible in the laboratories.  This possibility seems to be ruled out as there is no evidence for such an enhancement.
\vskip 0.3cm
$\bullet$ {\bf T(opology) Scenario}: In approaching the vector manifestation (VM) fixed point in  $bs$HLS  implemented with topology change,  $m_\rho^*\sim g_\rho^*\to 0$ (VM and scale-invariance) but $m_N^*\sim m_0\propto f_\sigma^*\neq 0$ (by scale symmetry), hence ${\cal M}=m_\rho^*/m_N^*\to 0$. I suggest that this scenario is favored in nuclear physics.

\section{Questions and Answers}
Let me now turn to what Nature says about the issues discussed above and what one can infer from them.
\subsection{Tensor forces}
The cusp observed above in the symmetry energy given by the skyrmion crystal can be reproduced by the behavior of the tensor forces in nuclear interactions.  It is due to the fact that nuclear symmetry energy is dominated by the tensor forces. In terms of the Lagrangian used in this paper, $bs$HLS, the (total) tensor force is given by one-pion and one-$\rho$ exchanges. In calculating the effective forces, one is to take into account the coupling constants and masses of the mesons involved in the exchanges that are scaling with density. The density dependence is prescribed by both the intrinsic density dependence (IDD for short) inherited from QCD at the scale at which the correlators of QCD and effective theory are matched and the in-medium renormalization group properties of the effective field theory, $bs$HLS.  

\begin{figure}[hb]
\begin{center}
\includegraphics[height=5.5cm]{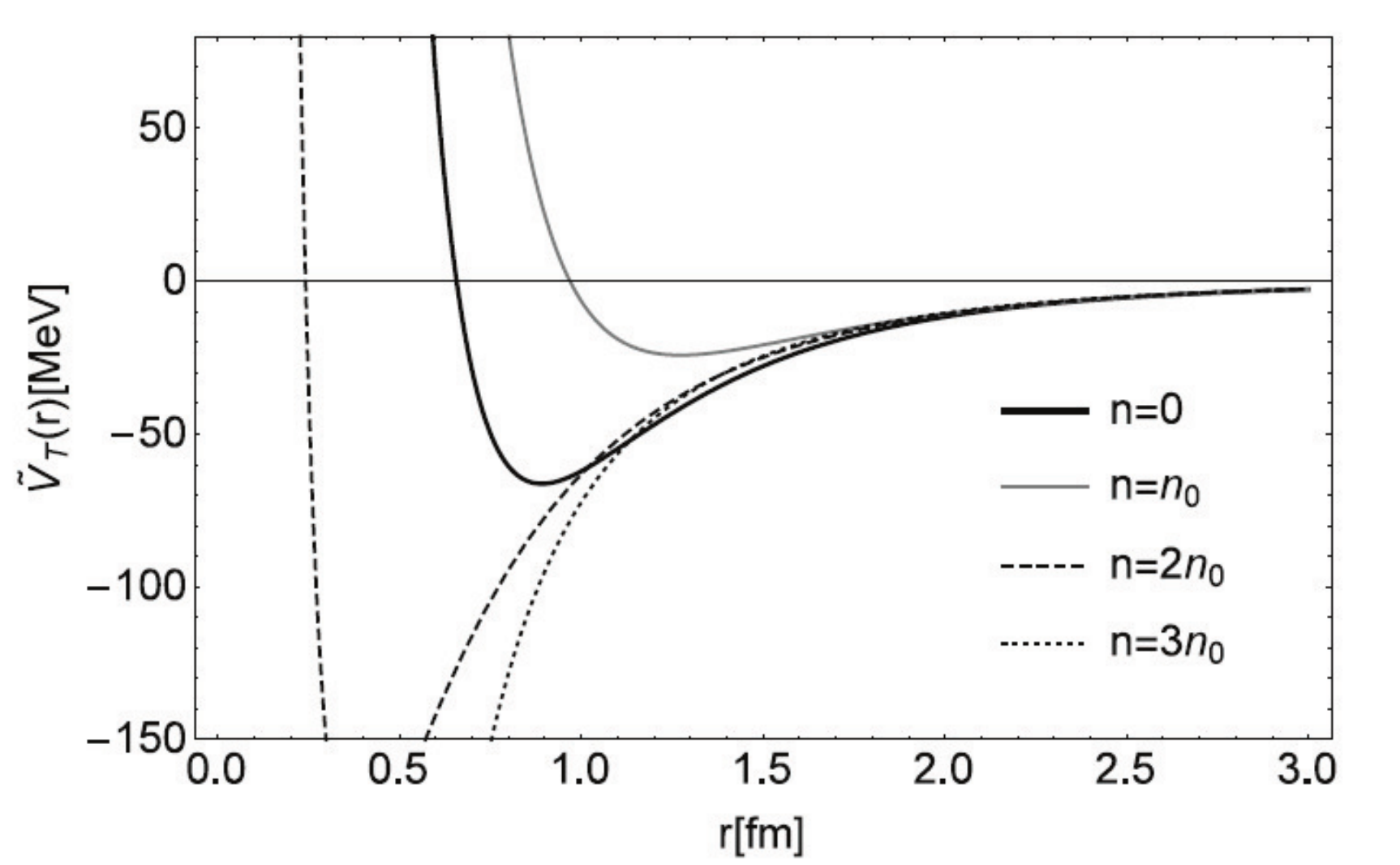}\includegraphics[height=5.9cm]{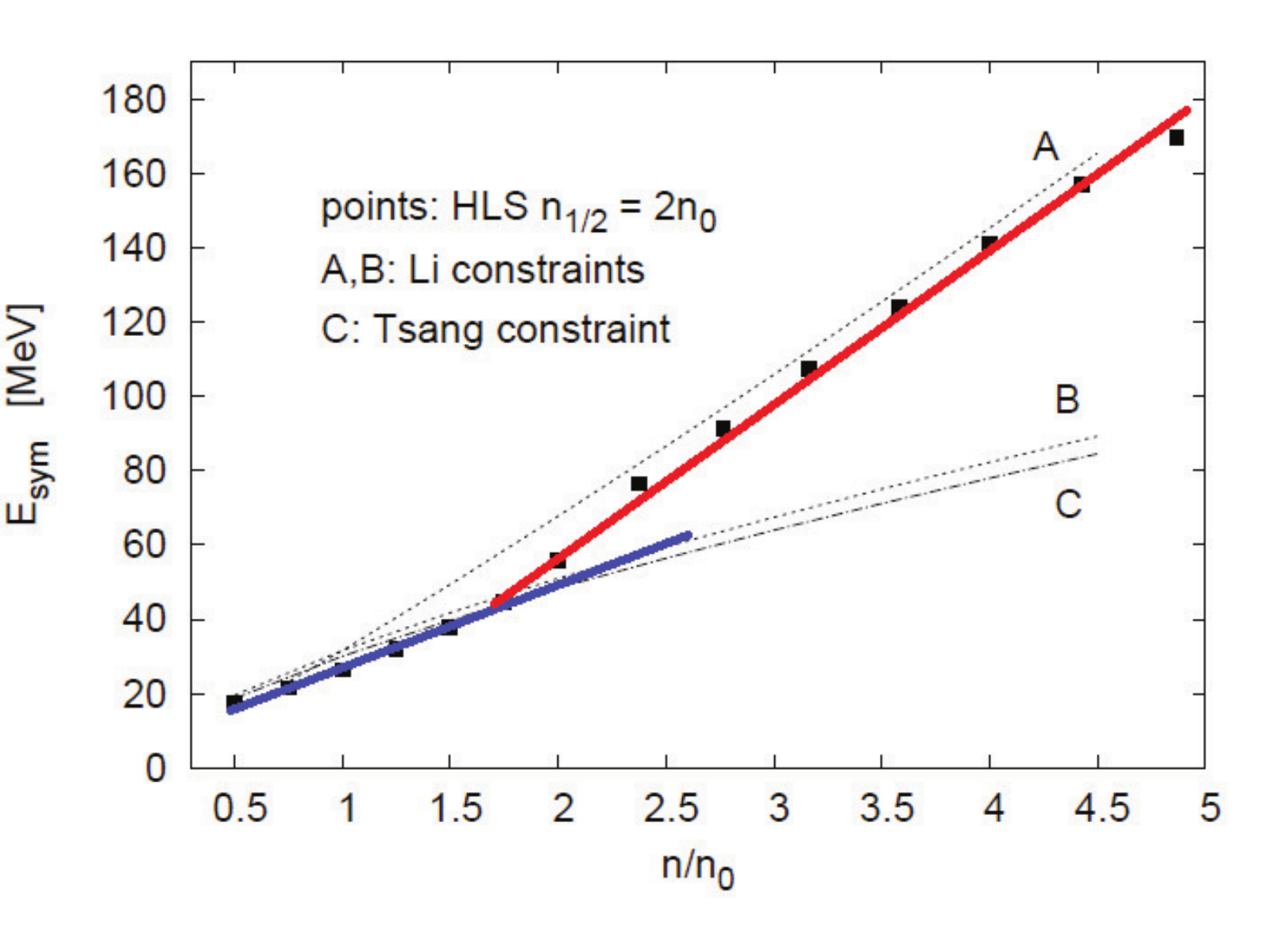}
\caption{Left panel: Sum of $\pi$ and $\rho$ tensor forces vs. $\mathbf{r}$ (fm) for different denstity $n$.  The $\rho$ tensor is completely suppressed at $n\gsim  2n_0$. Right panel: The symmetry energy $S$ predicted by the $V_{lowk}$ with $bs$HLS with $n_{1/2}=2n_0$. The eye-ball slope change is indicated by the colored straight lines..}
\label{tensor}
\end{center}
\end{figure}

What's involved here are two things. First the $\pi$-exchange tensor force and the $\rho$-exchange tensor force come with opposite signs so they tend to cancel in the range of forces entering in the pertinent nuclear interactions. The pion mass and coupling constants are protected by nearly exact chiral symmetry, consequently the pion tensor is more or less unaffected by density. On the other hand, the $\rho$ tensor increases in strength with the $\rho$ mass falling  as density increases toward $n_{1/2}$ and then gets quenched strongly by the hidden gauge coupling $g_\rho^*$ going toward zero as the VM fixed point is approached. The  net effect is then that the sum of the two tensor forces decreases due to the cancelation between them  up to $n_{1/2}$ and then is taken over by the pion tensor with the $\rho$ tensor suppressed, hence increasing beyond $n_{1/2}$. This property is seen in the left panel of Fig.\ref{tensor}.

Now the net effect of the tensor forces on the symmetry energy in nuclear matter near $n_0$ can be approximately expressed in the form that comes from the closure approximation $S \sim c\la |V^T|^2\ra/\bar{E}$ where $c$ is a dimensionless constant. This simple form follows from the fact that the tensor force predominantly excites  from the ground state to excited states peaked at $\bar{E}\sim (200-300)$ MeV.  This expression leads to the $S$ that decreases as $n$ approaches $n_{1/2}$ and then increases as $n$ goes above $n_{1/2}$. This is precisely the cusp structure predicted by the skyrmion model. It should be mentioned that this behavior of $S$ supports the scaling behavior of the $bs$HLS Lagrangian described above in the tensor force channel. It is important to note that what plays a crucial role here is the VM fixed point toward which dense matter flows.

\subsection{Probing the tensor force in nuclei}
It is fairly well recognized that the symmetry energy $S$ is the most important ingredient in the equation of state (EoS) for massive compact stars.  And it is also the most poorly known at high density. Thus the tensor force is the key ingredient for compact-star physics yet to be figured out. If the scenario described above is correct, a reliable approach to the EoS involves an intricate combination of topology change and in-medium properties of nucleons and mesons, specially the $\rho$ meson.

It should of course be recognized that the symmetry energy $S$ in nuclear matter is more involved than just what the tensor force gives. It can get contributions from different components of the nuclear force. In fact at low density, the closure-approximated expression for $S$ cannot be reliable. This is because first of all, the topology structure inferred from skyrmion crystal cannot be trusted at low density. Furthermore it is known that the cluster structure of matter can be important at low density. Nonetheless  the behavior for $n\gsim n_{1/2}$ is expected to be trustful as it is mainly controlled by the VM fixed point. A full RG flow calculation using $V_{lowk}$ potential described below gives the symmetry energy given in Fig.\ref{tensor} (right pannel). The cusp structure, smeared by other components of the nuclear force,  gives rise to a distinctive changeover from a soft $S$ to a stiffer $S$ at $n_{1/2}$. This stiffening accounts for $\sim 2$ solar mass compact star recently observed.

Exploring the behavior of the tensor force in nuclei is a fertile and challenging field in nuclear physics. I give a few illustrations as to where it stands.

\subsubsection{\it Carbon 14 dating}
The long lifetime for the Gamow-Teller transition in $^{14}$C presents a case where the structure of the tensor force is exposed at a matter density near $n_0$\cite{c14}. It sheds light on complex and highly intricate interplay between the pion tensor force and the $\rho$ tensor force at matter densities near $n_0$.  As shown in Fig.\ref{tensor} (left panel) and discussed above, up to $n_{1/2}\approx 2n_0$, the net tensor force at the range relevant to the process decreases as density increases, nearly vanishing at $\sim n_{1/2}$. This cancelation manifests spectacularly in the Gamow-Teller matrix element involved in the transition $^{14}$C $\to$ $^{14}$N because the GT matrix element nearly vanishes as density approaches $n_0$ around which the wave functions overlap. What is significant in this process is that the transition involved samples the density regime where such cancelation takes place and the energy spectrum in $^{14}$N for the corresponding states comes out in perfect agreement with experiments. This provides a consistency check between the weak current and the Hamiltonian, both of which scale with $f_\sigma^*=\la\chi\ra^*$ which, in scale-chiral effective theory, is locked  to $\la\bar{q}q\ra^*$.

It should be pointed out clearly that the cancelation in the transition matrix element, although consistent in a nontrivial way with the Hamiltonian giving the energy spectrum, involves an interplay of complicated many-body correlations and cannot single out unequivocally the role of the tensor force subject to the scaling parameters.\footnote{There are other more or less equally successful ways of obtaining the drastic suppression of the GT matrix element, such as, for example, by invoking three-body forces  instead of scaling parameters in, say, {\it ab initio} no-core shell-model approach. This does not imply however that they invalidate or replace the mechanism based on the scaling of \cite{c14}.   That basically they are more or less equivalent but described in different languages is explained in \cite{holt-rho-weise}.}
\subsubsection{\it Tensor force as a fixed-point interaction}
One must admit that the C-14 dating is at best a tenuous evidence for the effect one is looking for. One would like to zero-in on processes that directly exhibit the scaling behavior of the tensor force as a function of density. This will require precision in both theory and experiments. It would be mandatory to first search for experiments that can be done and analyzed with precision. This is a challenge for theorists and experimentalists.

From the theory side, there is one promising development, both surprising and intriguing. It is that the tensor force effective in nuclear interactions appears to be a fixed-point interaction in the sense of Landau Fermi-liquid fixed point theory. To be practical and concise  it is clearest to put it in terms of the $V_{lowk}$ potential\cite{vlowk} in the renormalization-group approach to effective theory for nuclear systems. In the Wilsonian RG approach, one can adopt what's called ``double-decimation RG"\cite{BR:DD}. $V_{lowk}$ is the effective potential arrived at from the first decimation with a cutoff at $\Lambda\sim 2$ fm$^{-1}$. It is applicable for the vacuum as well as in medium, endowed with suitable density dependence as described above. The second decimation brings it to the Landau Fermi-liquid fixed point and correlations on top of the Fermi surface.

It is not proven but observed in numerical results that the tensor force appears to be a fixed-point interaction. To clarify what's involved, let's think of approaching nuclear many-body problems using $V_{lowk}$ potentials constructed from $bs$HLS Lagrangian. This is a natural approach anchored on nuclear effective field theory exploiting RG decimations. Focusing on the tensor force, I denote by $V^{T} (n)$ the tensor potential given by the exchanges of $\pi$ and $\rho$ implemented with IDD defined at a scale relevant to nuclear calculations, roughly about the vector meson mass. In the framework adopted, this potential is applicable both in the vacuum and in medium. In the vacuum it will be $V^T(0)$. Denote by $V^T_{lowk} (n)$ the low-momentum component potential obtained from the first decimation defined with the cutoff $\Lambda\sim 2$  fm$^{-1}$ which seems to work well, corresponding approximately to the momentum scale up to which precision data are available for scattering data.

\begin{figure}[h]
 \begin{center}
 { \includegraphics[width=7cm,clip,angle=0]{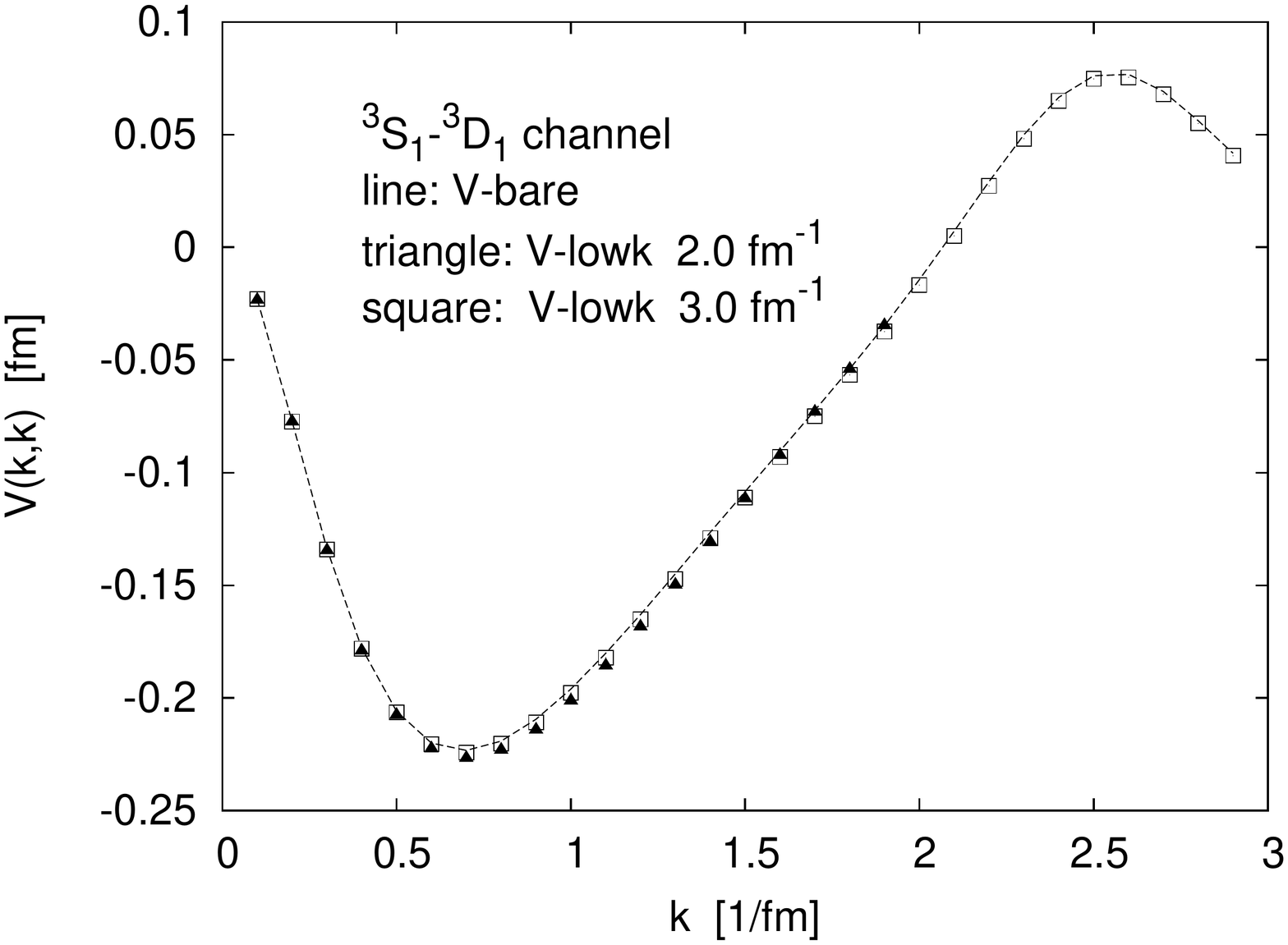} \includegraphics[width=7cm,clip,angle=0]{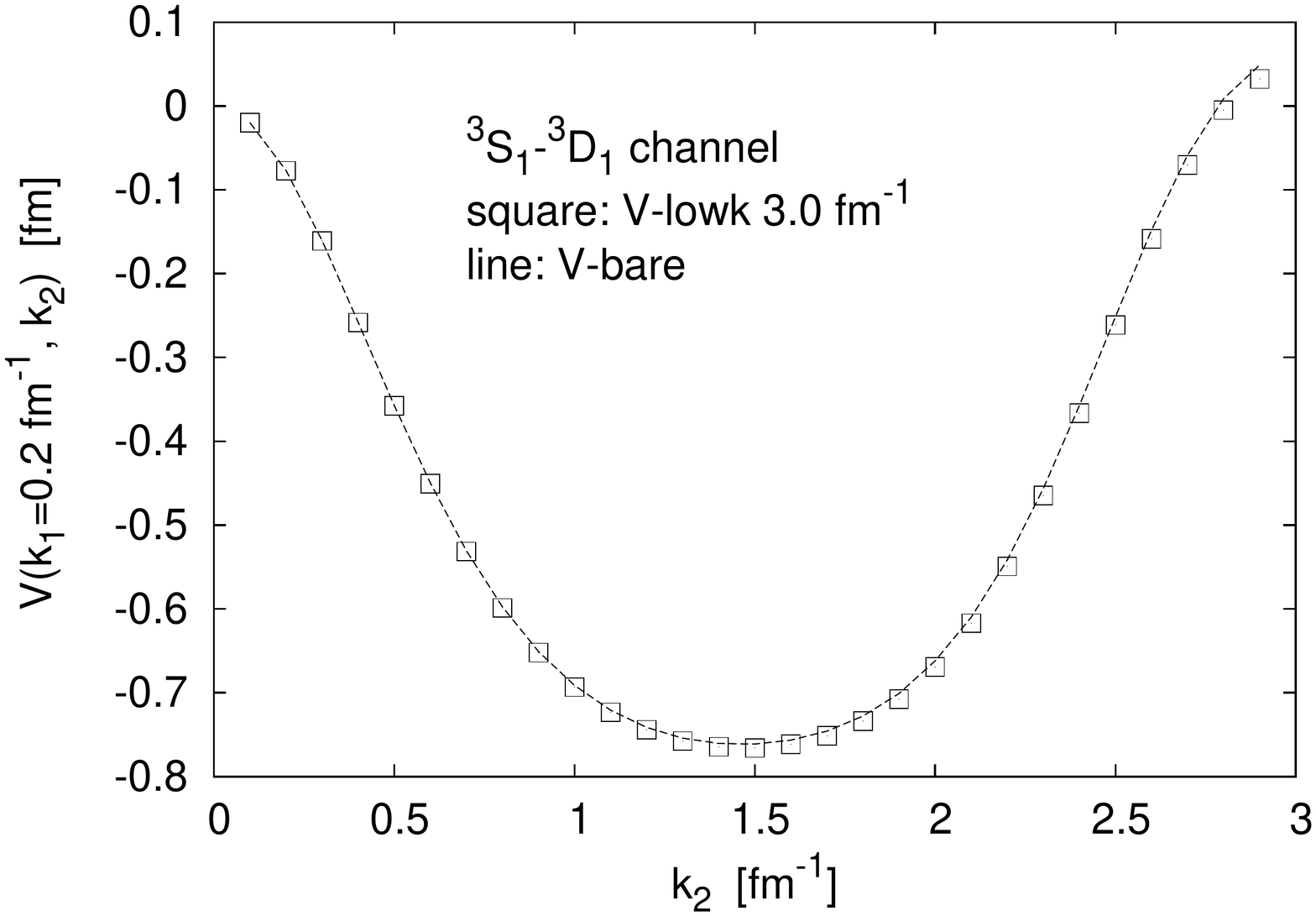}}
 \caption{``Bare" $V^T$ and $V^T_{low k}$ in matter-free space. Here the potential is the BonnS potential, approximating the leading-order contribution from the $bs$HLS Lagrangian.}
  \label{tensor-free}
 \end{center}
\end{figure}

The first observation is made in the matter-free space, $n=0$.  For this, no separation into double decimations is needed. Consider the two-nucleon matrix element in the tensor channel $^3{\rm S}_1$-$^3{\rm D}_1$ with $V^T$ and $V^T_{lowk}$. The surprising result shown in Fig.\ref{tensor-free}\footnote{These figures were obtained by Tom Kuo and sent to me in private communication for which I am deeply grateful.} is that
\be
\la f |V^T_{lowk}|i \ra\approx  \la f |V^T|i \ra
\ee
and the beta function
\be
\beta (V^T_{lowk})\approx 0.
\ee
This result suggests that the tensor force given at the matching scale is unrenormalized by strong interactions in the RG flow in matter-free space. Why the tensor force is at its fixed point, with the $\beta$ function nearly vanishing, has at present no explanation.

Let me go even further and look at processes in nuclei with density $n\lsim n_0$, which requires the double decimation including higher-order correlations. The C-14 dating case offers no clear hint due to the intricate interplay between the tensor force with other components of the nuclear force.  Fortunately there is a case, known up to date, relevant to the issue which hints at the direction to take. It is shell evolution in exotic nuclei.

It has been suggested that one of the underlying causes for shell evolution is based on the strong monopole part of effective nucleon-nucleon interactions for which the  tensor force is found to  play an important role in the appearance or disappearance of the spherical magic numbers in light-mass nuclei, and for heavier nuclei,  in the observed drift of single- particle levels along the $N= 51$ isotonic and the Sb ($Z= 51$) isotopic chains\cite{otsuka}.  The ``monopole" matrix element of the two-body interaction between two single-particle states labeled $j$ and $j^\prime$ and total two-particle isospin $T$ is given by  $V_{j,j^\prime}^T= \frac{\sum_J (2J+1)\la jj^\prime|V|jj^\prime\ra_{JT}}{\sum_J(2J+1)}$.
Remarkably this matrix element affects the evolution of single-particle energy;
$\Delta\epsilon_p (j)=\frac 12(V_{jj^\prime}^{T=0} +V_{jj^\prime}^{T=1})n_n(j^\prime)$
where  $\Delta\epsilon_p (j)$ represents the change of the single-particle energy of protons in the state $j$ when  $n_n(j^\prime)$ neutrons occupy the state $j^\prime$. It turns out that the matrix elements $V_{jj^\prime}$ and $V_{j^\prime j}$ have opposite sign for the tensor forces if $j$ and $j^\prime$ are spin-orbit partners.

\begin{figure}[htb]
\begin{center}
  \includegraphics[width=8cm,clip]{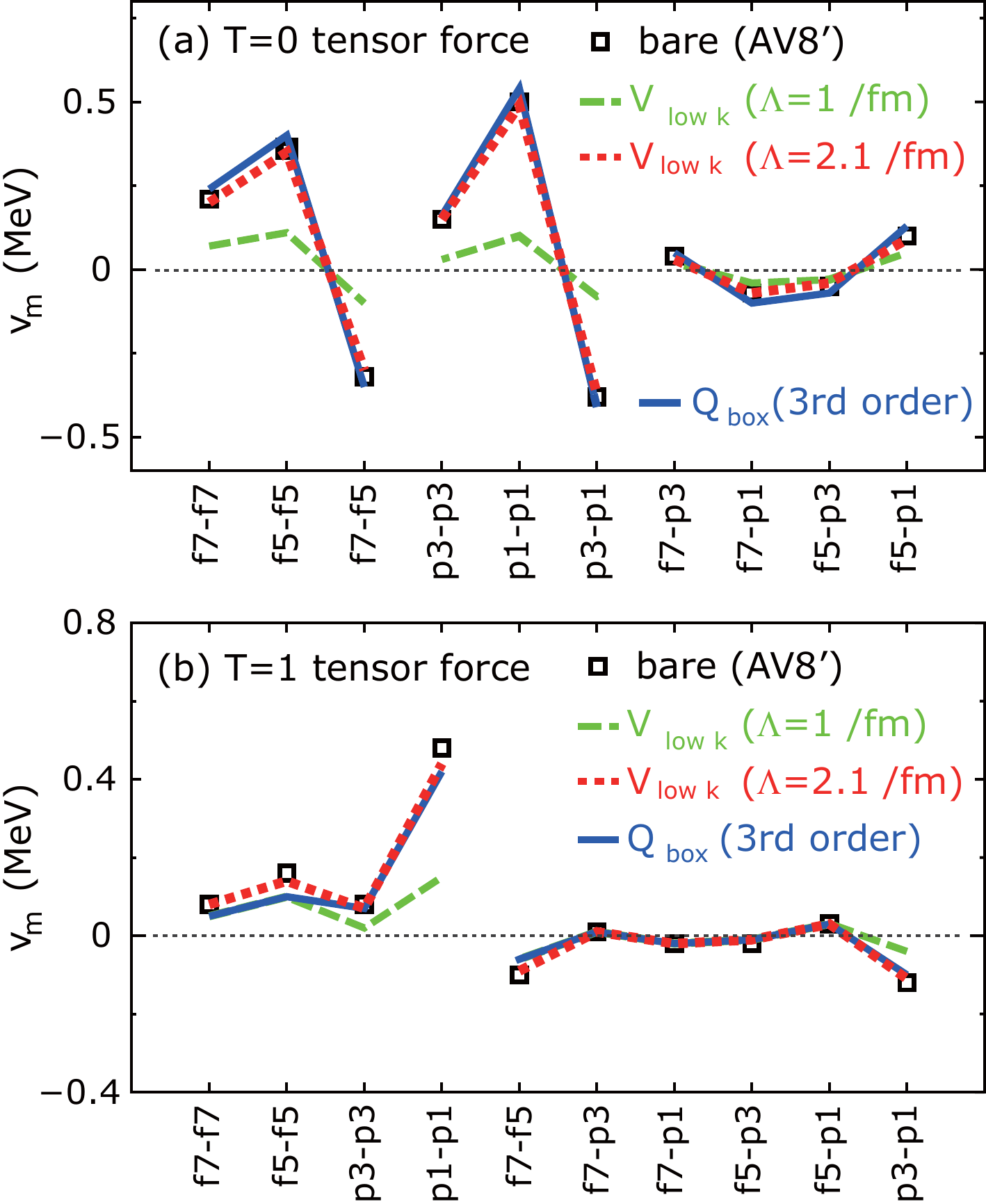}
 \caption{Tensor forces in AV8' interaction, in low-momentum interactions in the pf shell
obtained from AV8', and in the 3rd-order Q$_{box}$ interaction for (a) T=0 and (b) T=1. From \cite{otsuka}.}
  \label{pfshell}
 \end{center}
\end{figure}

Let me summarize the salient features of the shell evolution connected to the tensor force found by Otsuka\cite{otsuka,otsuka-persistence}.
The phenomenological potential Av18' and the one given by ChPT at N$^3$LO, both treated \`a la  $V_{\rm lowk}$ -- and also other realistic potentials  (such as BonnS) -- are found to give the same results. The cutoff ${\Lambda}$ is varied and the result is found to be independent of the cutoff around 2.1 ${\rm fm}^{-1}$.
Shown in Fig.\ref{pfshell} is the shell evolution in the pf and sd regions calculated  by including high-order correlations using the  Q-box formalism to 3rd order\cite{otsuka}. While the central part of the potential is strongly renormalized by high-order terms, the tensor forces are left unrenormalized, leaving the ``bare" tensors more or less intact.
What the result shows is that the sum of the short-range correlation and medium effects as taken into account by the 3rd order Q-box\cite{vlowk}  leaves the bare tensor force unchanged. Given the robustness of the result,  one may take this result to imply  that up to nuclear matter density
\be
\frac{d}{d\tilde{\Lambda}} V_{low k}^{tensor}=\beta([V_{low k}^{tensor}], \tilde{\Lambda}) \approx 0.\label{beta}
\ee
This is a surprisingly intriguing result. If one were to take this result seriously, it would mean that the tensor force is unrenormalized both in the free space and in medium. The Q-box used here corresponds to high-order correlation effects in nuclear medium, thus we arrive at the observation that highly involved nuclear correlations also cancel in the tensor force.

However the analysis made in \cite{otsuka,otsuka-persistence} contains no IDD, hence does not contain the information on the decrease of the tensor force in density  up to $n\sim n_0$ which plays an important role in the C14 dating. There are no experimental data yet to check whether the results in \cite{otsuka} confirm or rule out the density dependence in the tensor force predicted.  The fixed-point property, if valid, should apply with the IDD incorporated. It is a challenging problem to confront the IDD-implemented tensor force in the monopole matrix elements figuring in the shell evolution. This would require both highly sophisticated precision calculations and precision experimental data.

The scaling of the mass ratio  ${\cal M}\equiv m_\rho^*/m_N^*$  is expected to set in as the density  surpasses $n_{1/2}\sim 2n_0$ and  becomes more prominent as it goes toward the VM fixed point.  This will be an indication for  difference in the sources for the masses of baryons and  mesons. The RIB machines, both in operation and also in project,  could reach density $\gsim 2n_0$ and hence probe the regime where the changeover of the state of matter takes place. What can be studied in RIB and going beyond to what could be offered by more powerful machines like FAIR of Darmstadt will confront what will be observed in gravity waves coming from the merging of neutron stars.  The gravity waves  are estimated to provide information of density of order $\sim 6n_0$ at temperatures of order $\sim 60$ MeV\cite{shibata}. The onset of the mass ratio $\cal{M}$ triggers the change from soft to hard in the symmetry energy $S$ in compact stars as illustrated in Fig.\ref{tensor}. It will be extremely interesting to observe such changeover in the tidal deformability in compact stars as observable in GW. This possibility is a hot topic in nuclear astrophysics\cite{HKL-lovenumber}.

\section{Conclusion}
I discussed two possible scenarios for the origin of masses of light-quark hadrons, specifically the proton and the $\rho$ meson. As the baryonic matter density is increased beyond $n\sim 2n_0$ -- but before the particle picture breaks down, by e.g., possible percolation transition, the mass ratio ${\cal M}\equiv m_\rho^*/m_N^*$ could behave in different ways, signaling how the hadrons  shed their masses:
\begin{itemize}
\item {\bf ``Q mechanism"}: ${\cal M}\to  \sim 2/3$.  Going toward the chiral restoration density $n_c$, $m_\rho^*\to 0$ and $m_N^*\to 0$. This follows from the Nambu mechanism applied to quasiquarks with the assumption that the quasiparticle description continues to hold. Here hidden symmetries play no manifest role. 
\item {\bf ``T mechanism"}:  ${\cal M}\to  0$. Going toward the VM fixed point $n_{VM}$ -- which may or may not coincide with $n_c$, $m_\rho^*\to 0$ and $m_N^*\to f^*_\sigma\neq 0$. The former comes about because the $\rho$ mass is nearly scale-invariant\cite{yamawaki-GEB} and governed by the VM fixed point and the latter is controlled by (broken) scale symmetry.  In both,  the Cheshire Cat mechanism with  topology change, hidden symmetries -- with parity doubling and the vector manifestation -- and the special structure of the nuclear tensor forces play an essential  role.  The emergence of scale symmetry, hidden in nature, could take place {\it after} the VM fixed point and approaching the DLFP.  Deconfinement does not figure, at least up to compact-star-matter densities.
\end{itemize}
It is the latter scenario that, I suggest, is favored by a variety of nuclear phenomena treated in this note and is suggested for a future experimental discovery potential in nuclear physics.

On the theory side, there are  quite a few questions to be answered. Among them, regarding the half-skyrmion  on crystal lattice, could it be related to the chiral bag at the magic angle at which the baryon charge fractionizes into two halves? If so, how is it related to the topology change that plays an intriguing role for the tensor force and ultimately in the EoS for dense matter? Another issue is a possible Cheshire Cat interpretation of the hadron-quarkyonic transition\cite{quarkyonic} which resembles the skyrmion-half-skyrmion transition discussed in this note and detailed in\cite{PKLR}.  Finally the topology change could provide the additional attraction needed in the Akaishi-Yamazaki mechanism for high density kaonic-proton matter\cite{Akaishi-Yamazaki}. In \cite{Akaishi-Yamazaki}, the additional attraction was attributed to what was called ``vacuum clearing" in Ref \cite{vacuum-clearing}. In \cite{Park-Kim-Rho}, it was found that the changeover from skyrmions to half-skyrmions on crystal lattice gave rise to a substantial attraction over and above the Wess-Zumino term -- which is equivalent to the Weinberg-Tomozawa term used in \cite{Akaishi-Yamazaki} -- coming from the dilaton associated with the emergent scale symmetry. The scale-invariant hidden local symmetry scheme developed above could give a reliable value for this attraction.

\subsection*{Acknowldegments}
I would like to thank Youngman Kim for inviting me to write this note. I am very grateful to all my collaborators for extensive discussions on what's treated in this note.


\end{document}